# Classical Music Generation in Distinct Dastgahs with AlimNet ACGAN

Saber Malekzadeh
Computer Science Department
University of Tabriz
Tabriz, Iran
Saber.Malekzadeh@sru.ac.ir

Shahla Rezazadeh Azar
Computer Engineering Department
Khajeh-Nasir University of Technology
Tehran, Iran
ShahlaRezazadeh@email.kntu.ac.ir

Maryam Samami
Computer Engineering Department
Islamic Azad University, Sari branch
Sari, Iran
maryamsamami2013@gmail.com

Maryam Rayegan
Computer Engineering Department
Islamic Azad University, Shiraz branch
Shiraz, Iran
0smalek0@gmail.com

*Abstract*— **In this paper AlimNet (With respect to great musician, Alim Qasimov) an auxiliary generative adversarial deep neural network (ACGAN) for generating music categorically, is used. This proposed network is a conditional ACGAN to condition the generation process on music tracks which has a hybrid architecture, composing of different kind of layers of neural networks. The employed music dataset is MICM which contains 1137 music samples (506 violins and 631 straw) with seven types of classical music Dastgah labels. To extract both temporal and spectral features, Short-Time Fourier Transform (STFT) is applied to convert input audio signals from time domain to time-frequency domain. GANs are composed of a generator for generating new samples and a discriminator to help generator making better samples. Samples in time-frequency domain are used to train discriminator in fourteen classes (seven Dastgahs and two instruments). The outputs of the conditional ACGAN are also artificial music samples in those mentioned scales in time-frequency domain. Then the output of the generator is transformed by Inverse STFT (ISTFT). Finally, randomly ten generated music samples (five violin and five straw samples) are given to ten musicians to rate how exact the samples are and the overall result was 76.5%.**

## I. Introduction

Music is a complex sequential type of data. It appears at various timescales ranging from the periodicity of the waveforms at the scale of milliseconds, all the way to the musical form of a piece of music that may take several minutes. Music has a hierarchical structure, including a phrase which made up of smaller recurrent patterns (e.g., a bar). People pay attention to structural patterns related to coherence, rhythm, tension and the emotion flow [1]. The first model was built in order to combining algorithms in 1959 [2]. From 1989, shallow neural networks have applied for composing algorithms to music generation. Using shallow networks continued until recent years when deep neural networks presented their quality and capability in big data area that composing music using deep network has become popular. Most studies have been conducted on the music generation modeling using deep neural networks over the last three years [3].

Recurrent neural networks (RNNs) with long short-term memory (LSTM) cells have illustrated great results both in generating natural language and hand writing fields. For instance, RNN was used to generate a clean voice. Thereafter, to create vocals a deep RNN was proposed which in a discrimination was trained [4, 5]. The most well-known instances namely the MelodyRNN models and SimpleRNN models [6] which are specified for symbolic-domain and audio-domain generation respectively. So far, less studies have been devoted to use deep convolutional neural networks (CNNs) for creating music comparing to RNNs [3]. Not only Sample RNN, but also Wavenet have been applied for audio-domain generation. The major ingredient of WaveNet are causal convolutions. Since Wavenet and the models comprised with causal convolutions do not contain recurrent layers, they are faster to train compare to RNNs, especially when operated on long sequences data [7]. RNN and CNN are combined by adopting the Convolutional Recurrent Neural Network (CRNN) to model the audio features and achieved the state-of-art performance [8-10]. A Generative Adversarial Net (GAN) composed of two neural networks namely discriminator and Generator. GAN applies a discriminator network to train a generative model which has fulfilled the dreams of generating real-valued data. The Generator gains a random noise vector z to return an output which is the discriminator input [11]. To generate music in this paper, the Auxiliary Classifier GAN (ACGAN) is applied which was proposed in [12]. The model was created with assigning an additional structure with specialized cost function to the GAN [8]. The ACGAN function is specialized to separating big data sets into some subsets by class and training a Generator and discriminator for each subset [12]. The aforementioned model [12] is a kind of the GAN architecture in which Generator conditions output on its class label, and the discriminator implements auxiliary classification to recognize the fake and the real sample regarding to their respective class labels [13].

In this paper The proposed ACGAN includes a generator which is a deep neural network (DNN) to generate music from noise and also a discriminator which includes a hybrid architecture combining RNNs and CNNs aiming to be taught from music samples which are fed into Discriminator as a time-frequency domain sample [8].

## II. RELATED WORK

Authors in [9, 10] combined RNN and CNN by adopting the Convolutional Recurrent Neural Network (CRNN) to model the audio features and achieved the state-of-art performance [8].

Recently, some deep neural network models have been provided in order to generate a melody sequence or audio waveforms through a few priming notes and also in some cases combining a melody sequence with some other parts of music [6, 14-18].

One of the most famous symbolic-domain music generation models is named Melody RNN models. Generally, there are three RNN-based models namely the loopback RNN, the attention RNN and two types of RNN that aim to learn longer-term structures [3].

Song from PI [19] applies a hierarchy of recurrent layers in order to creating a multitrack song by generating melody, the drums and chords. This model is capable of generating several different sequences simultaneously. It is worth noting that the aforementioned model needs prior knowledge related to the musical scale to generate melody which is not needed in our applied model [19].

C-RNN-GAN takes random noises as input of Generator leading generates several kinds of music, though the model does not apply a conditional mechanism to generate music in its structure [20-22].

DeepMind provided a CNN-based model namely WaveNet which is considered as an audio-domain model. The model is probabilistic conditional one which is able to generate raw waveforms of speech and music and also has some advantages as follow: generating novel musical fragments, giving promising results about phoneme recognition [17, 23].

The MidiNet is a generative model which was proposed as a symbolic domain model. The model applies CNNs for generating melody in the form of the series of MIDI notes. More over a discriminator was used to learn the distributions of melodies. The model uses a new conditional mechanism to exploit prior knowledge to generate melodies not only from scratch, but also by conditioning on the melody of previous bars between other several possibilities [3].

The proposed TACGAN model, presents a generation model which in the input vector of the Generator is a noise vector z and also other vector comprising an embedded representation of the textual description. However, the applied discriminator is as the same as the ACGAN discriminator and also is augmented to achieve the text knowledge as input before classifying. Instead of assigning the class label to which the combined image is supposed to be fake, the noise vector $z_c^{\hat{}}$, including information regarding to the textual description of the image would be the input [13].

## III. THE PERLIMINARY

In this preliminary section, at first the applied dataset is described in detail. Secondly, the Short-time Fourier transform (STFT) and the inverse STFT (ISTFT) are explained with the formula respectively. Then, the ACGAN structures are represented. At last, the DNN structure is mentioned.

### A. Maryam Iranian classical music data set (MICM)

The applied dataset namely Maryam Iranian classical music contains 1137 music samples which includes 506 music samples with the foreground violin instrument and also some other instruments in the background. It is worth noting that the rest music samples use the Ney instrument as the foreground instrument. The reason behind applying two musical instrument, Violin and straw, is to provide an instrument Independent method to generating distinct Dastgahs[1] in Iranian traditional music data set. The given dataset has seven classes which represents the names of Iranian traditional music Dastgahs namely Shour, Homayoun Mahour, Segah, Chahargah, Rastpanjgah and Nava. In the Table1, the number of music samples existed in each class are illustrated as bellow. Each music samples contains different numbers of signal samples. It is worth noting that the sample rate of each music sample is 8192.

The Table 1 illustrates the numbers of music samples in each class.

TABLE I. TABLE 1. MICM SAMPLES DESCRIPTION

| Name of dastgah | number of music samples |
|---|---|
| Shour | 445 |
| Homayoun | 173 |
| Mahour | 150 |
| Segah | 74 |
| Chahargah | 106 |
| Rastpanjgah | 94 |
| Nava | 95 |

### B. Short-time Fourier transform (STFT)

To extract the frequency features of the audio signal, Fourier transform is used by many researchers. Fourier analysis has some disadvantages e.g., not being able to reflect the local time-domain information. Short-Time Fourier Transform (STFT) is used in this paper to extract the necessity information from the audio signals. STFT splits the signal into small time blocks; after that, it employs the Fourier transform to each time block [24].

---

[1] - Dastgah is a traditional Persian musical modal system which is a melody type.

$$F(\omega) = \int_{-\infty}^{+\infty} f(t) \exp(-i\omega t)\, dt. \quad (1)$$

where $i = \sqrt{-1}$.

The STFT formula for the time domain signal f(t) is shown as follows:

$$F_{STFT}(\tau,\omega) = \int_{-\infty}^{+\infty} f(t)\, g^*(t-\tau) \exp(-i\omega t)\, dt. \quad (2)$$

In the above formula, $\tau$ is the time shift parameter, the signal g(t) illustrates a fixed length window and the symbol (∗) provides the complex conjugate [25].

### C. The inverse STFT (ISTFT)

The output of the Generator should be converted to time-domain signals. Inverse STFT (ISTFT) is applied to reconstructing time-domain signals from their STFT without additional time-varying normalization [26].

Time-domain output signal $y_i(t)$ is computed by using an inverse STFT (ISTFT) formula as below [27]:

$$y_i(r+r) = \frac{1}{L.win(r)} \sum_{f \in \{0, \frac{1}{L} f_s, \ldots, \frac{L-1}{L} f_s\}} y_i(f,r) e^{j2\pi f r}. \quad (3)$$

### D. ACGAN architecture

The architecture of the ACGAN consists of a Generator and a Discriminator. Generator adopts deep neural network to generate music which is generated as a musical waveform in the audio domain, aiming fools the Discriminator. A Discriminator which applies deep neural networks to be able to recognize between the real and fake (generated) data, gives us an output which is close to 1 for real data (i.e. X) and 0 for the fake samples (i.e. G(z)). Let X be a dataset used for training the GAN and $I_{real}$ denotes a sample from X. Usually in GANs, Generator gives a vector of random noises $z \in R^L$, whereas it returns X = G(z) that seems to be real to Discriminator. But in in the ACGAN, every generated sample has a related class label, $c \sim p_c$ in addition to the noise z. Generator uses both to generate artificial data $X_{fake}$ = G(c,z). The discriminator not only returns a probability distribution over sources (fake and labels) but also gives the probability distribution over the class labels, DS(I) = P (S | I) and DC(I) = P (C | I). The objective function includes two sections: the log-likelihood of the correct source, LS, and the log-likelihood of the correct class, LC.

$$LS = E[\log P(S = real\,|X_{real})] + E[\log P(S = fake|X_{fake})]. \quad (5)$$

$$LC = E[\log P(C = c\,|X_{real})] + E[\log P(C = c|X_{fake})]. \quad (6)$$

During training Discriminator try to maximize LS + LC while the aim of a Generator is minimizing LC −LS [12].

### E. Dnn structure

A DNN is a feed-forward neural network that generally contains more than one layer of hidden neuron among the input the output layer.

CNNs are one particular type of deep, feed forward network composed of kernels that have learnable weights. Each kernel convolves on an input data and activation function is applied to the given convolution result. A CNN is a kind of score function as receives a STFT sample on one end to output class scores at the other end. CNNs also contain a loss function to calculate the cost of the network prediction to be idol and optimizing results by reforming the weights in back propagation operation with an optimizer function.

In the proposed deep model, Gated Recurrent unit (GRU) a new kind of RNN layer is used. A RNN is a type of artificial neural network where connections among neurons create a directed graph along a sequence. RNNs employ their memory to process sequences of inputs. RNN, relates all the sequences of inputs together. In the prediction or generation cases, the relation among all the previous words or samples helps in predicting or generating the better result. The RNN produces the networks with loops in them, causing to persist the information.

## IV. THE PROPOSED METHOD

In the proposed method section, the processing stages applied to the given data set are described. Then, the process of STFT application on the preprocessed data and also the output result is shown in detail. At last, the proposed method is explained in detail with the structure.

### A. Preprocessing steps of MCIM music sample

As known, DNNs just are able to get data samples with the same length as inputs. As mentioned in the previous sub-section, each sound sample included in the MICM has different lengths. Each sound sample is cut. The all cut music samples have 131072 signal samples. As previously mentioned the sample rate of each music sample is 8192. Therefore, each sound sample contains 16 seconds of music. The reason behind this selection is because, Dastgahs in Iranian Classical music can be recognized easily with 16 seconds of music.

### B. STFT application on the preprocessed data

In order to have a time-frequency domain data, STFT is applied to the preprocessed data which is in time domain. STFT contains some input parameters which change the output of the STFT e.g., changing the size and resolution of the output. One of these parameters is fast Fourier transform (FFT) windows size which is set to 510 in this paper. The next parameter named hop length which represent the number of the frames of audio between STFT columns.

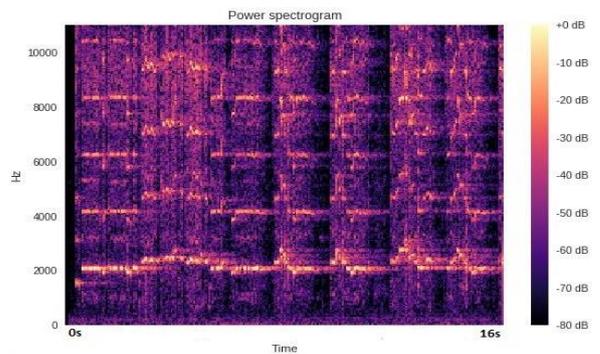

Figure 1. Scaled STFT music samples

The mentioned parameter is set to 514. Applying the given parameters to the preprocessed music signal leads to achieve an output with 256*256 matrix. The provided output matrix illustrates both spectral and temporal features of the input audio signal in time-frequency domain. The figure 1 represents the output STFT sample, but in input samples of AlimNet samples are not scaled between 0 and -80.

*C. The proposed method with the audio Representation*

In the proposed method, every produced music is associated with a class label c and a random vector z, which is typically from a uniform distribution or normal distribution. The class label with noise vector are used as the input of the generator to generate music tracks. The output of the generator is a music that is used as the input of the discriminator which distinguish the real samples from generated ones. In the output layer of discriminator, a sigmoid function is used to return output results in the range of [0,1]. The discriminator is optimized with a cross-entropy loss function, to drive the output of discriminator to 1 for real data (i.e. X) and 0 for the fake (i.e. G(z)). The generator tries to create outputs close to the real data in the given scales in order to fool the Discriminator [28]. To train the discriminator, STFT samples are fed in to it as inputs. The mentioned variant of GANs applies label conditioning that outputs several music tracks in fourteen different classical music Dastgahs. The proposed ACGAN in this paper is conditioned on the class label and the discriminator not only is able to distinguish real STFT samples from the generated ones, but also is able to determine a correct class label to each sample. Worth to note that the input noise size for the generator was a 1*256 matrix.

Although the convolutional layer is used to recognition of local conjunctions of features, such as extracting the features, from the layer below [29], gated recurrent unit (GRU) is applied to temporal summarization of the extracted features [30]. GRU is used in this paper as it is able to make each recurrent unit to take dependencies of different time scale [30].

In practice, the applied Generator and the Discriminator are DNN and GRU respectively, with the following architecture shown in fig.2 and fig.3 respectively. The architecture of discriminator as a classifier is more likely to the proposed Azarnet DNN in our previous paper [31].

The architecture of the discriminator is shown in Table 2.

| Layer type | Output shape | # Parameters |
| --- | --- | --- |
| 2D Convolution (3*3)(16) | (256, 256, 16) | 160 |
| Dropout (0.1) | (256, 256, 16) | 0 |
| Batch Normalization (0.8) | (256, 256, 16) | 64 |
| 2D Max Pooling (2*2) | (128, 128, 16) | 0 |
| 2D Convolution (3*3)(32) | (128, 128, 32) | 4640 |
| Dropout (0.2) | (128, 128, 32) | 0 |
| Batch Normalization (0.8) | (128, 128, 32) | 128 |
| 2D Max Pooling (2*2) | (64, 64, 32) | 0 |
| 2D Convolution (3*3)(32) | (64, 64, 32) | 9248 |
| Dropout (0.3) | (64, 64, 32) | 0 |
| Batch Normalization (0.8) | (64, 64, 32) | 128 |
| 2D Max Pooling (2*2) | (32, 32, 32) | 0 |
| 2D Convolution (3*3)(32) | (32, 32, 32) | 9248 |
| Dropout (0.3) | (32, 32, 32) | 0 |
| Batch Normalization (0.8) | (32, 32, 32) | 128 |
| 2D Max Pooling (2*2) | (16, 16, 32) | 0 |
| 2D Convolution (3*3)(64) | (16, 16, 64) | 18496 |
| Dropout (0.4) | (16, 16, 64) | 0 |
| Batch Normalization (0.8) | (16, 16, 64) | 256 |
| 2D Max Pooling (2*2) | (8, 8, 64) | 0 |
| Reshape | (64, 64) | 0 |
| GRU (50) | (64, 50) | 17400 |
| GRU (100) | (100) | 45600 |
| FC (5) | (5) | 505 |
| FC (7) (classifier) | (7) | 42 |

The architecture of the generator is shown in Table 3.

| Layer type | Output shape | # Parameters |
| --- | --- | --- |
| FC | (256) | 65792 |
| Reshape | (16, 16, 1) | 0 |
| Batch Normalization (0.8) | (16, 16, 1) | 64 |
| UpSampling2D (2*2) | (32, 32, 1) | 0 |
| 2D Convolution (3*3)(256) | (32, 32, 256) | 70452 |
| Batch Normalization (0.8) | (32, 32, 256) | 256 |
| UpSampling2D (2*2) | (64, 64, 256) | 0 |
| 2D Convolution (3*3)(128) | (64, 64, 128) | 35894 |
| Batch Normalization (0.8) | (64, 64, 128) | 128 |
| UpSampling2D (2*2) | (128, 128, 128) | 0 |
| 2D Convolution (3*3)(64) | (128, 128, 64) | 18496 |
| Batch Normalization (0.8) | (128, 128, 64) | 64 |
| UpSampling2D (2*2) | (256, 256, 64) | 0 |
| 2D Convolution (3*3)(32) | (256, 256, 32) | 9248 |
| Dropout (0.3) | (32, 32, 32) | 0 |
| Batch Normalization (0.8) | (32, 32, 32) | 128 |
| 2D Max Pooling (2*2) | (16, 16, 32) | 0 |
| 2D Convolution (3*3)(64) | (16, 16, 64) | 18496 |

## V. CONCLUSION

By conditioning the input of generator on the given class labels, the Conditional ACGAN is able to generate samples regarding to the intended classes. The outputs of the conditional ACGAN are also artificial music samples in those mentioned scales in time-frequency domain. Then the output of the generator is transformed by Inverse STFT (ISTFT). Finally, ten generated music samples (five violin and five straw samples) are given to ten musicians randomly to rate the quality of the generated samples and the overall result was 76.5%.

## REFERENCES

[1] 1. Dong, H.-W., et al. "MuseGAN: Multi-track sequential generative adversarial networks for symbolic music generation and accompaniment". in Proc. AAAI Conf. Artificial Intelligence. 2018.

[2] 2. Hiller, L.A. and L.M. Isaacson, Experimental music: composition with an electronic computer. 1959.